\documentclass[
  aps,
  prl,
  twocolumn,
  superscriptaddress,
  nofootinbib
]{revtex4-2}

\usepackage{graphicx}
\usepackage{amsmath}
\usepackage{amsfonts}
\usepackage{mathrsfs}
\usepackage{bbm}
\usepackage{xcolor}
\usepackage[colorlinks=true]{hyperref}
\usepackage{ulem}
\usepackage{braket}
\usepackage[T1]{fontenc}

\begin{document}

\title{Thermal responses of spin density and spin current induced by Bogoliubov Fermi surfaces in proximity superconductors}	

\author{Yingwen Zhang}
\affiliation{Department of Physics, City University of Hong Kong, Kowloon, Hong Kong SAR, China}

\author{Cong Xiao}
\email{congxiao@fudan.edu.cn}
\affiliation{Interdisciplinary Center for Theoretical Physics and Information Sciences (ICTPIS), Fudan University, Shanghai 200433, China}
\affiliation{State Key Laboratory of Surface Physics, Fudan University, Shanghai 200433, China}

\author{Xiao Li}
\email{xiao.li@cityu.edu.hk}
\affiliation{Department of Physics, City University of Hong Kong, Kowloon, Hong Kong SAR, China}

\begin{abstract}

The emergence of Bogoliubov Fermi surfaces in $s$-wave proximity superconducting systems provides a unique platform for generating nontrivial spin responses in the presence of spin-orbit coupling. In this work, we investigate temperature-gradient driven responses of spin polarization and spin current in such a proximity system related to Fe(Te,Se). The interplay of in-plane magnetization, superconductivity, and nematicity gives rise to nontrivial Berry curvature and spin-related hybrid Berry curvature on the Bogoliubov Fermi surfaces, which are shown to lead to finite thermally driven nonequilibrium spin polarization and spin Nernst effect. We find that the responses are strongly enhanced by the emergence of Bogoliubov Fermi surfaces, and are linear in temperature at low temperatures, rendering a unique spintronic character of Bogoliubov Fermi surfaces. 

\end{abstract}


\maketitle

\textit{Introduction—}In conventional $s$-wave superconductors, the Bogoliubov quasiparticle spectrum is fully gapped at low energies. Under certain conditions, however, this gap can close, giving rise to gapless superconducting phases characterized by Bogoliubov Fermi surfaces and a finite quasiparticle density of states at zero energy \cite{agterberg2017bogoliubov,Sato2017,brydon2018bogoliubov}. Such gapless structures can originate from nonunitary interband pairing enabled by additional internal degrees of freedom, such as orbital or sublattice indices \cite{gao2010interorbital,Nomoto2016multiSC,cheung2019superconductivity,menke2019bogoliubov}, or emerge in noncentrosymmetric systems with strong spin–orbit coupling when superconductivity is induced via proximity effects in the presence of an in-plane Zeeman field \cite{xu2015structured,yuan2018zeeman,zhu2021discovery}.

Upon the emergence of Bogoliubov Fermi surfaces, the nontrivial quantum geometric properties of quasiparticle wave functions can be manifested in low-energy responses driven by, e.g., temperature gradient. In particular, when inversion symmetry is broken and spin–orbit coupling is present, the Bogoliubov Fermi surface is expected to exhibit not only momentum-space Berry curvature but also the Berry curvature defined in hybrid momentum-magnetization space \cite{yuriy2017,xiong2018,Yuriy2019mixed, Dong2020}. In normal-state spintronics, such a hybrid Berry curvature has been known to play crucial roles in nonequilibrium spin polarization phenomena \cite{freimuth2014, Freimuth2015Onsager,yuriy2017,xiong2018,Yuriy2019mixed, Dong2020,xiao2021sot,xiao2023,tang2024lossless,Manchon2024pumping}. This naturally motivates the investigation of Berry phase effects tied to Bogoliubov Fermi surfaces in spin-related thermoelectric responses of superconducting systems.

The iron-based superconductor Fe(Te,Se) serves as a unique intrinsic platform that combines superconductivity, Dirac surface states, and ferromagnetism. Recent experimental observation have confirmed the coexistence of these features at the interface \cite{qiu2023emergent}. The coexistence of proximity superconductivity and in-plane Zeeman field at such a interface enables the formation of Bogoliubov Fermi surfaces even in the absence of external magnetic field, providing a promising platform for exploring nontrivial spin responses due to Berry curvatures properties of Bogoliubov Fermi surfaces.

In this work, we investigate temperature-gradient induced spin polarization and spin transport in proximity-induced superconducting systems related to
iron-based superconductor Fe(Te,Se).
To describe the low-energy physics, we first consider a minimal monolayer model, and then incorporate electronic nematicity, a widely observed feature of iron-based superconductors arising from orbital inequivalence \cite{johnson2015spin,zhang2016distinctive,yi2017role,long2020first,lohani2020band,mascot2022topological,morfoot2023resurgence,wu2024nematic}. 
We show that the interplay of in-plane magnetization, superconductivity, and nematicity can lead to the formation of Bogoliubov Fermi surfaces with nontrivial momentum-space Berry curvature and spin-related hybrid Berry curvature. 
We combine symmetry analysis and model calculations to understand the spin polarization and spin current thermal responses, with particular emphasis on the role of Bogoliubov Fermi surfaces. We find that the responses are strongly enhanced by the emergence of Bogoliubov Fermi surfaces, and are linear in temperature at low temperatures, providing a unique spintronic character of Bogoliubov Fermi surfaces. 

\textit{Response formulas—}In the semiclassical theory, in the presence of a temperature gradient $\nabla T$, the local spin density up to the first order of spatial gradient is given by \cite{Dong2020} 
\begin{equation}
  \begin{aligned}
        \delta {\boldsymbol{s}}_{\mathrm{loc}}
             = -\partial_i M^{i\boldsymbol{s}}-\chi^{i\boldsymbol{s}}\partial_iT,
  \end{aligned} \label{eq:9}
\end{equation}
where $\partial_i \equiv \partial/\partial r_i$, and the summation over repeated Cartesian indices is implied. Here, the index $i$ denotes the spatial direction, while $s$ labels the spin component. The first term on the right hand side is a total derivative and vanishes after bulk average, whereas the second term is proportional to the temperature gradient and gives rise to the pursued averaged macroscopic response.
As both terms are unfamiliar in the context of superconducting response, here we explain briefly their physical meaning.

In the first term, 
\begin{equation}
M^{i\boldsymbol{s}}=\int(fm^{i\boldsymbol{s}}+g\Omega_{k_i}^{\boldsymbol{s}})
\end{equation}
is the dipole density of spin \cite{Dong2020}, or called spin part of the magnetic quadrupole density \cite{Gao2018spin,Shitade2019spin}. The equilibrium spin dipole density itself is an important physical quantity usually considered in the field of multiferroics, but has not been studied in superconducting systems. The notation $\int$ without integral variable is shorthand for $\sum_n\int d^D\boldsymbol{k}/(2\pi)^D$, where $\boldsymbol{k}$ and $n$ are respectively the momentum and band index of the quasiparticle, and $D$ is the dimensionality of the system. $f$ is the local equilibrium Fermi-Dirac distribution, and $g=k_BT\ln{(1-f)}$ is the grand potential density carried by each quasiparticle. $M^{i\boldsymbol{s}}$ is the lowest order multipole of spin \cite{xiao2010berry,Culcer2004,Xiao2021conserved} characterizing spin nonconservation due to spin-orbit coupling.
\begin{equation}
    m^{i\boldsymbol{s}} = \sum_{\ell \neq n}{\rm{Im}}\frac{\bra{u_n}\hat{v}_i\ket{u_\ell}\bra{u_\ell}\hat{\boldsymbol{s}}\ket{u_n}}{E_n-E_\ell}
\end{equation}
is the dipole moment of spin on each spin-orbit coupled quasiparticle state $\ket{u_n}$ with energy $E_n$.
Here, $\hat{v}_i = \partial_{k_i}\hat{H}_{\rm BdG}$ is the velocity operator in the Bogoliubov-de Gennes (BdG) framework, and $\hat{\boldsymbol{s}}$ is the spin operator.
For Bogoliubov quasiparticles, the spin operator can be written as $\hat{s}_x=\tau_z \sigma_x, \hat{s}_y=\tau_0 \sigma_y$, and $\hat{s}_z=\tau_z \sigma_z$ in Nambu space. $\boldsymbol{\tau}$ and $\boldsymbol{\sigma}$ are Pauli matrices acting on particle-hole space and spin space, respectively. The interband matrix element of $\hat{\boldsymbol{s}}$ being involved in the expression ensures mathematically that $m^{i\boldsymbol{s}}$ is a unique property of spin-orbit coupled quasiparticles. On the other hand,
\begin{equation}
    \Omega_{k_i}^{\boldsymbol{s}}=-2\sum_{\ell\neq n}{\rm{Im}}\frac{\bra{u_n}\hat{v}_i\ket{u_\ell}\bra{u_\ell}\hat{\boldsymbol{s}}\ket{u_n}}{(E_n-E_\ell)^2} \label{eq:11}
\end{equation}
is the spin counterpart of the usual momentum space Berry curvature, with one velocity operator replaced by the spin operator here. Similar quantities have also appeared in, e.g., layer-dipole quantum geometric response of magnetic bilayer systems, where one velocity operator in the numerator is replaced by the dipole operator and the resulting Berry-curvature-like quantity is called dipole Berry curvature \cite{fan2024intrinsic}. Here, following this convention, we term $\Omega_{k_i}^{\boldsymbol{s}}$ as the spin Berry curvature.

In the second term of Eq.~\eqref{eq:9}, the response coefficient $\chi^{i\boldsymbol{s}}$ of temperature gradient induced spin polarization is expressed as \cite{Dong2020}
\begin{equation}
    \begin{aligned}
        \chi^{i\boldsymbol{s}}=\frac{1}{T}\int (f E-g) \Omega_{k_i}^{\boldsymbol{s}}
    \end{aligned} \label{eq:10}
\end{equation}
for Bogoliubov quasiparticles. The appearance of spin Berry curvature in both $\chi^{i\boldsymbol{s}}$ and $M^{i\boldsymbol{s}}$ is not a coincidence, but manifests the theoretical structure of a generalized Streda formula
\begin{equation}
    \begin{aligned}
        \chi^{i\boldsymbol{s}}=-\frac{\partial M^{i\boldsymbol{s}}_{\mathrm{B}}}{\partial T}, \label{streda}
    \end{aligned} 
\end{equation}
where $M^{i\boldsymbol{s}}_{\mathrm{B}}=\int g\Omega_{k_i}^{\boldsymbol{s}}$ is the Berry-curvature part of the spin dipole density. This formula is parallel to the familiar Streda formula in thermoelectric transport of charge, where the thermoelectric Nernst conductivity is linked to the orbital magnetization \cite{xiao2006}, i.e., velocity dipole density, by the same formula \cite{Xiao2020EM}.

The formulation of spin current response is much more involved if we employ the conserved spin current definition \cite{Shi2006} to address the non-conserving issue in the conventional definition of spin current ($\hat{j}^s=\{\hat{s},\hat{v}\}/2$). This is because the conserved spin current does not have a well defined operator form compatible with periodic systems. Therefore, while direct evaluation of spin current response to temperature gradient has been done, the pertinent theory is highly complicated \cite{Xiao2021conserved,liao2025semiclassical}. On the other hand, the outcome of that theory is compact: The transport spin current reads as
\begin{equation}
\boldsymbol{\mathcal{J}}^{s_{\alpha}}=\boldsymbol{\alpha}_{\mathrm{H}}^{s_{\alpha}} \times \nabla T,
\end{equation}
where the spin Nernst conductivity is a vector in real space and is given by
\begin{equation}
    \boldsymbol{\alpha}_{\mathrm{H}}^{s_{\alpha}} =\frac{1}{T}\int (f E-g) s_{\alpha} \boldsymbol{\Omega}_{\boldsymbol{k}},
\end{equation}
with $ \boldsymbol{\Omega}_{\boldsymbol{k}}$ being the vector form of the usual momentum space Berry curvature \cite{xiao2010berry} and $s_\alpha=\bra{u_n}\hat{s}_\alpha\ket{u_n}$.
The simplicity of this result reminds that there should be a convenient way to approach it. Indeed, according to the Onsager reciprocity \cite{Shi2006,Xiao2021conserved}, it can be obtained by calculating the linear response of entropy current density to the gradient of a Zeeman field \cite{Xiao2020EM,Pan2024PRR}. Taking the Berry curvature anomalous velocity \cite{Sundaram1999}, this reciprocal response is given by
\begin{equation}
     \boldsymbol{j}= \int (-\boldsymbol{F}_{\boldsymbol{k}} \times \boldsymbol{\Omega}_{\boldsymbol{k}})\mathfrak{s},
\end{equation}
where \(\boldsymbol{F}_{\boldsymbol{k}}=-s_{\alpha}\nabla m_{\alpha}\) is the driving force due to the auxiliary Zeeman gradient $\nabla m_{\alpha}$, and \(\mathfrak{s}= \frac{1}{T}(fE-g)\) is the entropy density carried by each quasiparticle. One thus finds that the response of charge current $\boldsymbol{j}$ with respect to $\nabla m_{\alpha}$ is indeed given by nothing but $\boldsymbol{\alpha}_{\mathrm{H}}^{s_{\alpha}}$.

\textit{Model calculation and discussion—}Ferromagnetism and superconductivity are typically considered mutually exclusive orders, as they tend to align electron spins in different configurations. Notably, their coexistence has been experimentally observed in van der Waals Josephson junctions based on the iron-based superconductor Fe(Te,Se) \cite{qiu2023emergent}, where two Fe(Te,Se) flakes are separated by a van der Waals gap. Here, we consider the case of a monolayer on the surface of Fe(Te,Se), where the superconducting gap is induced from the bulk via the proximity effect. The Hamiltonian formulated in the BdG form in momentum space is given by
\begin{equation}
H=\frac{1}{2}\sum_{\boldsymbol{k}} \psi_{\boldsymbol{k}}^{\dagger}H_{\rm{BdG}}(\boldsymbol{k})\psi_{\boldsymbol{k}},\label{eq:12}
\end{equation}	
where we define the Nambu spinor operator $\psi_{\boldsymbol{k}}^{\dagger}=(c_{{\boldsymbol{k}}\uparrow}^{\dagger},c_{{\boldsymbol{k}}\downarrow}^{\dagger},c_{-{\boldsymbol{k}}\uparrow},c_{-{\boldsymbol{k}}\downarrow})$, with \(c_{{\boldsymbol{k}}\uparrow}^{\dagger}\) as the creation operator in a single layer.
The effective Hamiltonian describing the monolayer can be written as
\begin{equation}
H_{\rm{BdG}}=
    \begin{bmatrix}
        H_0(k_x,k_y) & \Delta_0\\
        -\Delta^*_0    & -H^*_0(-k_x,-k_y)
    \end{bmatrix}, \label{eq:13}
\end{equation}
where 
\begin{equation}
\begin{aligned}
     H_0 &= -\mu \sigma_0 + v_F (k_x \sigma_y-k_y \sigma_x)+V_x \sigma_x \label{eq:14}
\end{aligned}
\end{equation}
is a \(2\times 2\) matrix that describes the low-energy features of the Dirac surface states with an in-plane magnetization.
Here, \(\mu\) is the electron chemical potential of the surface state, $v_F$ denotes the Fermi velocity in the surface Dirac Hamiltonian and \(V_x\) is the in-plane spin magnetization. Besides, the off-diagonal terms in the BdG Hamiltonian~\eqref{eq:13} are \(2\times 2\) matrices with \(\Delta_0=i\Delta \sigma_y \).

\begin{figure}
    \centering
    \includegraphics[width=1\linewidth]{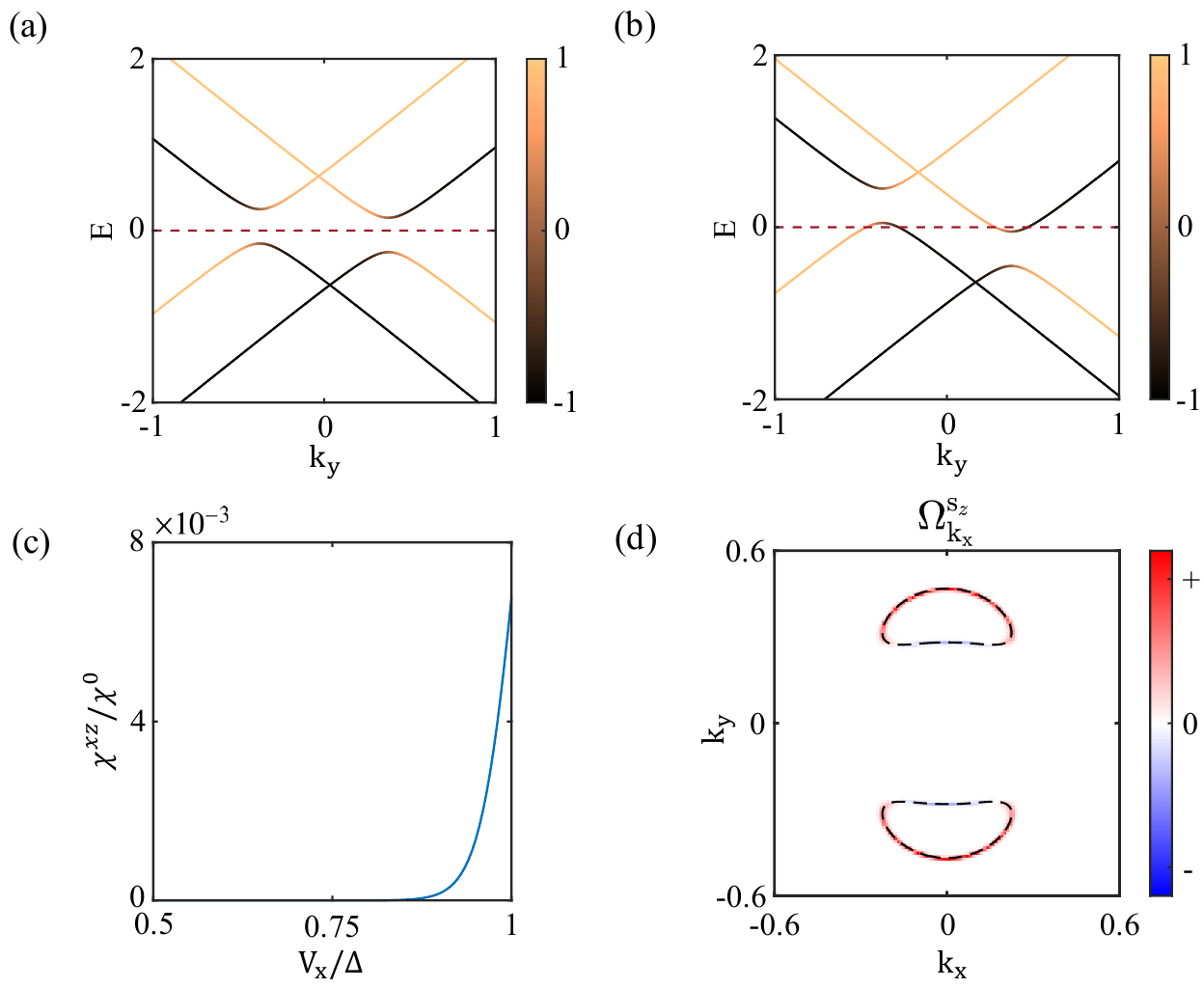}
    \caption{(a) Energy dispersion as a function of $k_y$ in the absence of Bogoliubov Fermi surfaces $V_x=0.5\Delta$. (b) Energy dispersion for $V_x=1.25\Delta$, where Bogoliubov Fermi surfaces emerge and the spectrum becomes gapless. The color scale represents the effective charge of the Bogoliubov quasiparticles. (c) Dependence of the spin polarization response coefficient $\chi^{xz}$ on $V_x$ at $T=0.01\Delta$. Here, $\chi^0=\hbar k_B/v_F$. (d) The distribution of spin Berry curvature $\Omega_{k_x}^{s_z}$ around the Bogoliubov Fermi surfaces (dashed lines), with the same parameters as in (b).}
    \label{fig:1}
\end{figure}

The Hamiltonian of the monolayer possesses the magnetic point group symmetry $m'm2'$, which is invariant under the symmetry operations \(\{M_x,M_yT,C_{2z}T\}\). According to the symmetry requirements listed in Table \ref{table:1}, $\chi^{xz}$ (i.e., $\chi^{xS_z}$) is the only one tensor element of $\chi$ allowed by symmetry. It means that when a temperature gradient is applied along the $x$ direction, a spin polarization along the $z$ direction is induced in the system.

\begin{table}[b]
        \setlength{\tabcolsep}{6.5 pt} 
	\caption{Transformations of temperature gradient and spin (\(s_x,s_y,s_z\)) under symmetry operations of model~\eqref{eq:13}, as well as symmetry allowed components of $\chi$. Here, "\(+\)" represents the term is invariant under symmetry
operations, while "$-$" means the term becomes the opposite.}	\centering
	\begin{tabular}{c c c c c c}
		\toprule
		 symmetries & \(\partial_x T\) & \(\partial_yT\) & \((s_x,s_y,s_z)\) & allowed $\chi^{i\boldsymbol{s}}$ \\
		\hline 
		\(M_x\) & $-$ & $+$ & $(+,-,-)$ & $(\chi^{xy},\chi^{xz},\chi^{yx})$\\
		\(M_yT\) & $+$ & $-$ & $(+,-,+)$  &$(\chi^{xx},\chi^{xz},\chi^{yy})$\\
            \(C_{2z}T\) & $-$ & $-$ &  $(+,+,-)$ &$(\chi^{xz},\chi^{yz})$ \\
		\toprule
	\end{tabular} \label{table:1}
\end{table}

In the monolayer model investigated here, the in-plane magnetization $V_x$ plays a critical role in the emergence of Bogoliubov Fermi surface and intrinsic thermal spin polarization response. When $V_x$ is finite, the bands remain symmetric with respect to $k_x=0$ and without exhibiting any tilt along the $k_x$ direction. This allows us to set $k_x=0$ and analyze the conditions when the band structure becomes gapless. By solving for the eigenvalues of the Hamiltonian Eq.~\eqref{eq:13} at $k_x=0$, we obtain the following expressions for the four bands,
\begin{equation}
    E_{\pm} = \pm \sqrt{\Delta^2 +(v_F k_y-\mu)^2} \pm V_x.
\end{equation}
It follows directly that when $V_x>\Delta$,  Bogoliubov Fermi surfaces emerge and the system transitions into a gapless phase.
The evolution of the quasiparticle spectrum is illustrated in Fig. \ref{fig:1} (a) and (b), which show the BdG band dispersions for representative values of $V_x<\Delta$ and $V_x>\Delta$, respectively. In the former case, a global superconducting gap is present. In contrast, when \(V_x\) exceeds the gap, the bands cross the Fermi level at zero energy, signaling the formation of Bogoliubov Fermi surfaces and the onset of a gapless superconducting state.

\begin{figure}
    \centering
    \includegraphics[width=1\linewidth]{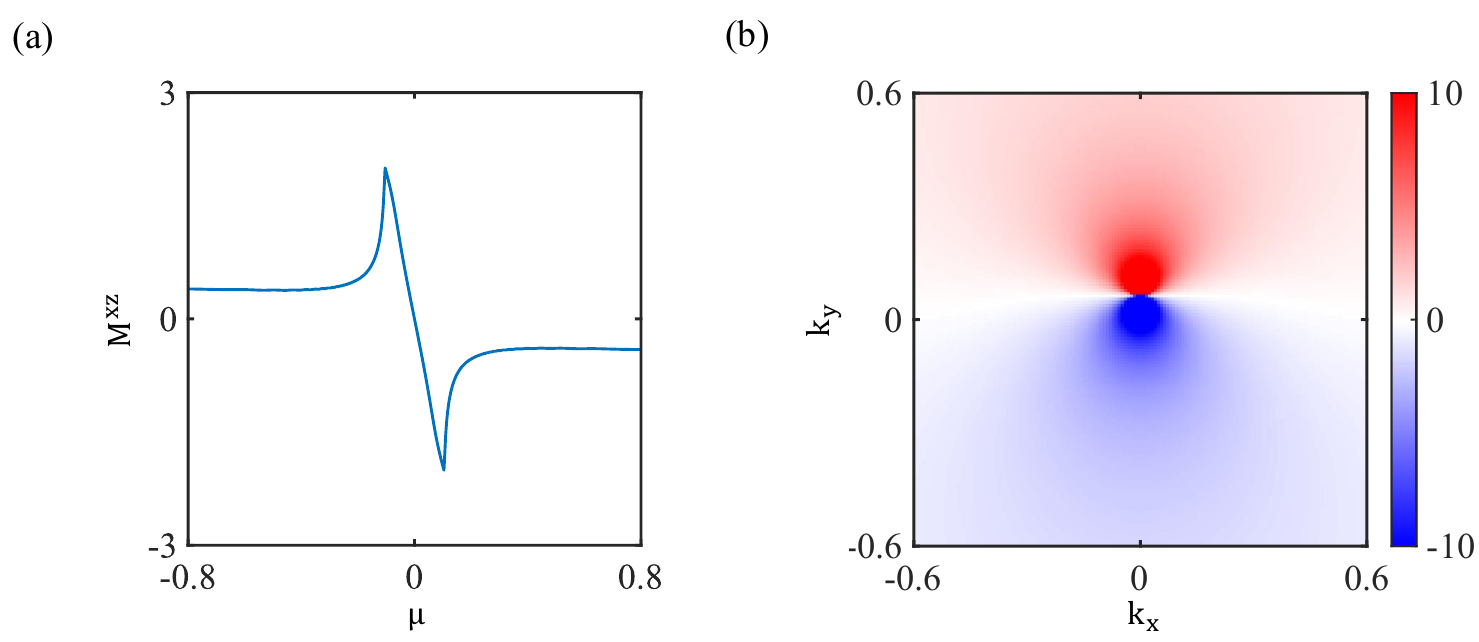}
    \caption{(a) Spin dipole density $M^{xz}$ as a function of the normal state chemical potential $\mu$. 
    (b) $k$-space distribution of the integrand of the spin dipole density, i.e., $f m^{xz} + g\Omega_{x}^z$ (in units of $v_F/\Delta$). In the calculations, we take $T = 0.03\Delta$ and $V_x=0.5\Delta$.}
    \label{fig:2}
\end{figure}

\begin{figure*}[t]
    \centering
    \includegraphics[width=\textwidth,height=0.9\textheight,keepaspectratio]{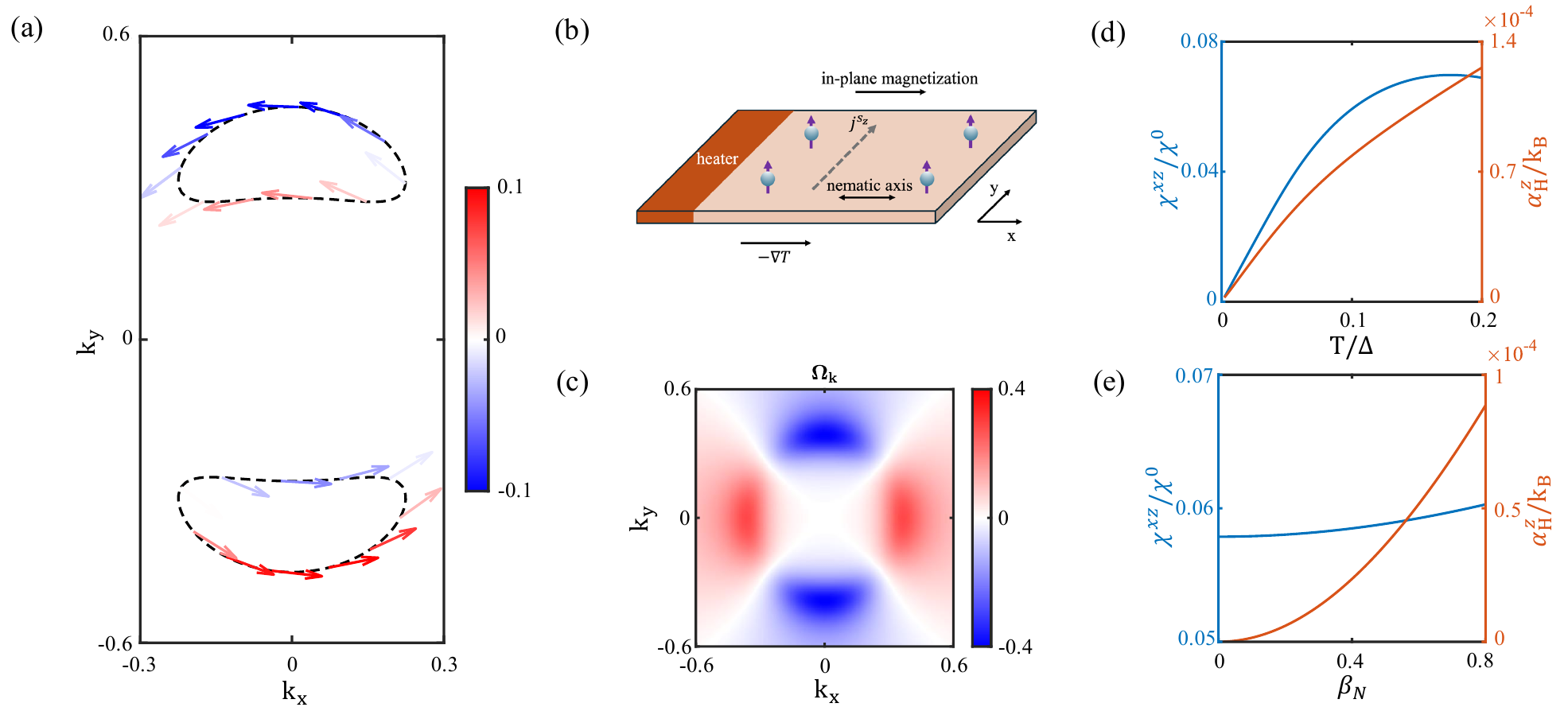}
    \caption{(a) Spin texture of the quasiparticles on the Bogoliubov Fermi surfaces (dashed line) in the presence of the nematic term. The arrows indicate the in-plane components of the spin, whereas the color scale represents the out-of-plane component $s_z$ (dimensionless). (b) Schematic illustration of the thermal response geometry. The temperature gradient $\nabla T$ and the in-plane magnetization $V_x$ are both applied along the $x$ direction. The double arrow indicates the orientation of the nematic axis. The thermally induced spin polarization is along the out-of-plane $z$ direction and thermally induced transverse spin current $j^{s_z}$ is along the $y$ direction. (c) Berry curvature distribution $\boldsymbol{\Omega}_{\boldsymbol{k}}$ in momentum space generated by the nematic correction for the positive-energy ($E>0$) bands. The color scale is in units of $(v_F/\Delta)^{2}$. (d) Thermal spin response coefficient \(\chi^{xz}\) (blue line, left axis) and spin Nernst conductivity \(\alpha_{\mathrm{H}}^{z}\) (orange line, right axis) as a function of temperature. In the calculations, we take $V_x=1.25\Delta$, $\mu=3\Delta$ and $\beta_N=3{v^2_F/\Delta}$. (e) Thermal spin response coefficient \(\chi^{xz}\) (blue line, left axis) and spin Nernst conductivity \(\alpha_{\mathrm{H}}^{z}\) (orange line, right axis) as a function of the strength of the nematic anisotropy. We take $T=0.1\Delta$ here.}
    \label{fig:3}
\end{figure*}

Figure \ref{fig:1}(c) displays the dependence of the spin thermal response coefficient \(\chi^{xz}\) on the in-plane magnetization \(V_x\) at a low temperature $T=0.01\Delta$. It is observed that the response becomes significant only when the system approaches the gapless phase with Bogoliubov Fermi surfaces. Consistently, in Fig. \ref{fig:1} (d), one finds significant distribution of spin Berry curvature on the Bogoliubov Fermi surfaces.

The appearance of equilibrium spin dipole density requires breaking both inversion and time reversal symmetries, which is the case here. In fact, according to the generalized Streda formula (\ref{streda}), $M^{i\boldsymbol{s}}$ and $\chi^{i\boldsymbol{s}}$ should possess the same symmetry requirements. Therefore, only one tensor element \(M^{xz}\) (i.e., $M^{xs_z}$, the dipole of spin-$z$ component in the $x$ direction) is allowed in the present model. Figure \ref{fig:2} (a) illustrates the dependence of the spin dipole density $M^{xz}$ on the normal state chemical potential $\mu$. A sharp change is observed near $\mu \approx 0$, where \(M^{xz}\) undergoes a sign change. This behavior is rooted in the microscopic distribution shown in Fig. \ref{fig:2} (b) where the integrand exhibits a prominent dipole-like distribution in momentum space. The center of this dipole pattern is shifted along the $k_y$ direction because the magnetization $V_x$ induces a shift of the Dirac point along this direction by $V_x/v_F$.

Although the minimal model in Eq.~\eqref{eq:13} captures the essential ingredients of the surface Dirac dispersion, the exchange field, and the superconducting proximity effect, it does not incorporate the orbital-related anisotropy in Fe(Te,Se). Experimental and theoretical studies have revealed substantial inequivalence between the \(d_{xz}\) and \(d_{yz}\) orbitals, accompanied by anisotropic reconstructions of the low-energy electronic structure \cite{johnson2015spin,lohani2020band,mascot2022topological,morfoot2023resurgence,wu2024nematic}. To phenomenologically incorporate this orbital anisotropy into the low-energy surface-state description, we introduce an additional nematic correction to the Hamiltonian. The leading symmetry-allowed correction in the momentum expansion takes the form of \cite{lohani2020band,morfoot2023resurgence}
\begin{equation}
    H_{Nem}=\beta_N (k_x^2-k_y^2)\sigma_z,
\end{equation}
where $\beta_N$ is the parameter describing the strength of the nematic anisotropy.

The nematic term introduces a momentum-dependent out-of-plane component in the spin texture in momentum space, leading to a finite $k$-space Berry curvature in the $z$ direction, which cannot be generated in model~\eqref{eq:13} without this nematic term. 
As illustrated in Fig. \ref{fig:3}(a), after adding the nematic term, the quasiparticles on the Bogoliubov Fermi surfaces develop a prominent out-of-plane spin component, as denoted by the color scale. 
The geometry of the thermal response is schematically illustrated in Fig. \ref{fig:3}(b).
The generated Berry curvature distribution is shown in Fig. \ref{fig:3}(c), which may induce spin Nernst transport.
We perform calculations on the spin density and spin current responses to temperature gradient, and find nonzero results, as shown in Fig. \ref{fig:3}(d). In particular, one finds that at low temperatures, both response coefficients exhibit a linear dependence on temperature. This linear-in-$T$ relation originates from the low-temperature expression of entropy density \(\mathfrak{s}= \frac{1}{T}(f E-g) \simeq \frac{1}{3}\pi^2k_B^2T\delta(E)\), which leads to
\begin{align}
    \chi^{i\boldsymbol{s}} & =\frac{1}{3}\pi^2k_B^2T\int \delta(E) \Omega_{k_i}^{\boldsymbol{s}}, \nonumber \\
    \boldsymbol{\alpha}_{\mathrm{H}}^{s_{\alpha}} & =\frac{1}{3}\pi^2k_B^2T\int \delta(E) s_{\alpha} \boldsymbol{\Omega}_{\boldsymbol{k}}.
\end{align}
One sees that the integral is nonzero only if there are Bogoliubov Fermi surfaces.
In contrast, when there is a global superconducting gap, both responses are significantly suppressed at low temperatures. The linear-in-$T$ dependence of the responses at low temperatures can thus serve as a hallmark of the Bogoliubov Fermi surface. Furthermore, we examine the dependence of these thermal responses on the nematic strength, as shown in Fig. \ref{fig:3}(e). As the nematic strength $\beta_N$ increases, both the spin Nernst response and the spin polarization are enhanced, while the spin Nernst response exhibits a significantly stronger dependence on $\beta_N$.

\textit{Summary and Discussion—}In summary, we have analyzed the thermally driven spin polarization and spin current responses in a model describing the proximity induced superconducting interface system related to Fe(Te,Se). The interplay of in-plane magnetization and proximity superconductivity generates Bogoliubov Fermi surfaces, whereas the spin-orbit coupling and nematic term shape the non-coplanar spin texture on the Bogoliubov Fermi surfaces, resulting in nonzero Berry curvature and spin Berry curvature. These ingredients can induce nontrivial nonequilibrium spin polarization and spin Nernst effect, as we show by model calculations. Our results revealed that the emergence of Bogoliubov Fermi surfaces leads to a pronounced enhancement of these responses at low temperatures. Notably, the linear-in-$T$ dependence serves as a distinctive signature of gapless Bogoliubov quasiparticles, providing a hallmark for such exotic superconducting states.

Although direct detection of a spin current remains experimentally challenging, nonlocal electrical detection based on spin-to-charge conversion, such as the inverse spin Hall effect \cite{sinova2015spin}, may provide one possible route to probe the thermally induced spin response predicted in this work. Magneto-optical techniques such as the Kerr effect \cite{sinova2015spin,Stamm2017Magneto} may provide another possible route, since they can be sensitive to the accompanying spin accumulation around the boundary of the sample.

The competition between intrinsic and extrinsic contributions to thermally induced superconducting spintronics effects remains an important open issue. In this work, we have focused on the intrinsic responses arising from the Berry curvature and spin Berry curvature of the Bogoliubov quasiparticle bands. On the other hand, a systematic theory of extrinsic thermal transport in superconducting systems is still lacking, and the scaling analysis commonly used in anomalous Hall transport is not directly applicable here because thermal response coefficients contain explicit and significant temperature dependence from the distribution function and thermal weighting factors, in addition to relaxation effects. 
Nevertheless, some qualitative insights from normal-state transport are expected to remain relevant. In particular, the skew scattering-induced extrinsic contribution is typically more pronounced in cleaner systems, while intrinsic and side-jump contributions are important in moderately dirty systems. Moreover, if the frequency of the driving field, here the temperature gradient, is comparable to or even exceeds the quasiparticle relaxation rate, scattering-induced extrinsic contributions, including the side-jump term, will be strongly suppressed, thereby allowing the intrinsic contribution to be isolated
\cite{nadvornik2021broadband,matsuda2023ultrafast,fujimoto2024observation}.

\textit{Acknowledgments—}Y.Z. thanks Zhi Wang and Zhen-Cheng Liao for the valuable discussions. X.L. is supported by the Research Grants Council of Hong Kong (Grants No. CityU 11304823, CityU11312825, C7012-21G, and C7015-24G) and City University of Hong Kong (Project No. 9610428).
C.X. is sponsored by National Natural Science Foundation of China (Grant No. 12574114) and the start-up funding from Fudan University.

\par\medskip
{\footnotesize
\noindent
\copyright\ 2026 Chinese Physical Society and
IOP Publishing Ltd. All rights, including
for text and data mining, AI training,
and similar technologies, are reserved.
\par}


\end{document}